\documentclass[preprint,superscriptaddress,showpacs,preprintnumbers,amsmath,amssymb,aps,pre]{revtex4}
\usepackage[utf8]{inputenc} %%%
\usepackage{epsfig}
\usepackage{natbib}
\usepackage{amsmath}
\usepackage{times}
\usepackage{subfigure}
\usepackage{epstopdf} %%%%
\usepackage{graphicx}
\usepackage{color}
\usepackage{pict2e}

\begin{document}

\title{Resonant behavior and unpredictability in forced chaotic scattering}

\author{Alexandre R. Nieto}
\email[]{alexandre.rodriguez@urjc.es}
\affiliation{Nonlinear Dynamics, Chaos and Complex Systems Group, Departamento de
F\'{i}sica, Universidad Rey Juan Carlos, Tulip\'{a}n s/n, 28933 M\'{o}stoles, Madrid, Spain}

\author{Jes\'{u}s M. Seoane}
\affiliation{Nonlinear Dynamics, Chaos and Complex Systems Group, Departamento de
F\'{i}sica, Universidad Rey Juan Carlos, Tulip\'{a}n s/n, 28933 M\'{o}stoles, Madrid, Spain}

\author{J. E. Alvarellos}
\affiliation{Departamento de F\'{i}sica Fundamental, Facultad de Ciencias, Universidad Nacional de Educaci\'{o}n a Distancia, Paseo Senda del Rey 9, 28040, Madrid, Spain}

\author{Miguel A. F. Sanju\'{a}n}
\affiliation{Nonlinear Dynamics, Chaos and Complex Systems Group, Departamento de
F\'{i}sica, Universidad Rey Juan Carlos, Tulip\'{a}n s/n, 28933 M\'{o}stoles, Madrid, Spain}
\affiliation{Institute for Physical Science and Technology, University of Maryland, College Park, Maryland 20742, USA}

\date{\today}

\begin{abstract}
Chaotic scattering in open Hamiltonian systems is a topic of
fundamental interest in physics, which has been mainly studied in
the purely conservative case. However, the effect of weak
perturbations in this kind of systems has been an important focus of
interest in the last decade. In a previous work, the authors studied
the effects of a periodic forcing in the decay law of the survival
probability, and they characterized the global properties of escape
dynamics. In the present paper, we add two important issues in the
effects of periodic forcing: the fractal dimension of the set of
singularities in the scattering function, and the unpredictability of
the exit basins, which is estimated by using the concept of basin
entropy. Both the fractal dimension and the basin entropy exhibit a
resonant-like decrease as the forcing frequency increases. We
provide a theoretical reasoning, which could justify this decreasing
in the fractality near the main resonant frequency, that appears for $\omega
\approx 1$. We attribute the decrease in the basin entropy to the
reduction of the area occupied by the KAM islands and the basin
boundaries when the frequency is close to the resonance. Finally,
the decay rate of the exponential decay law shows a minimum value of the
amplitude, $A_c$, which reflects the complete destruction of the KAM
islands in the resonance. We expect that this work could be potentially
useful in research fields related to chaotic Hamiltonian pumps,
oscillations in chemical reactions and companion galaxies, among
others.

\end{abstract}

\maketitle

%%%%%%%%%%%%%%%%%%%%%%%%%%%%%%%%%%%%%%%%%%%%%%%%%%%%%%%%%%%%%%%%%%%%%%%%%%%%%%%%%%%%%%%%%%%%%%%%%%%%%%%%%%%%%%%%%
\section{Introduction} \label{sec:Introduction}
Chaotic scattering in open Hamiltonian systems has been an interesting topic of
research in nonlinear dynamics and chaos theory due to its numerous applications in several fields of physics, such as celestial mechanics, fluid mechanics and atomic and nuclear physics (see Ref.~\cite{Seoane13} for an exhaustive description of the applications of chaotic scattering).
\\\indent In a generic situation of chaotic scattering, the particles
enter a finite region where they experience some sort of transient
chaotic dynamics due to the interaction with a potential, and then
leave the region. In this sense, the phenomenon could be understood
as a manifestation of transient chaos \cite {Lai, Tel}. Far enough
from the scattering region, the action of the potential is
negligible so the particle motion is essentially free. A usual tool
in chaotic scattering is the scattering function, which represents
the relation between characteristic variables of the input and the
output of the scattering region. When the scattering is nonchaotic,
the scattering function will be formed by smooth curves, which leads
to a high capacity to predict the behavior of one of the variables
from the other. In chaotic scattering problems, the scattering
function has a set of singularities as a result of the sensitive
dependence to the initial conditions, which constitutes a hallmark
of chaos. If we go into the neighborhood of one of the
singularities, we will conclude that the range of variation of the
output variable does not tend to zero as the size of the
neighborhood is arbitrarily reduced.
\\\indent Chaotic scattering processes have been studied in several physical systems such as hard-disk systems \cite{Gaspard} and billiard systems \cite{Stock,Graf}, although much of the literature refers to open Hamiltonian systems (e.g. \cite{Contopoulos92,Contopoulos93,Zotos17b,Kandrup99}), where the H\'{e}non-Heiles Hamiltonian constitutes a well-known paradigm (e.g. \cite{Aguirre01,Barrio08,Barrio09,Vallejo03,Zotos17}). The Hamiltonian is given by
\begin{equation} \label{eq:HH_Hamiltonian}
{\cal{H}}=\frac{1}{2}(\dot{x}^2+\dot{y}^2)+\frac{1}{2}(x^2+y^2)+x^2y-\frac{1}{3}y^3.
\end{equation}
\indent This system is conservative and hence the energy is conserved. Furthermore, it has a critical value of the energy $E_e=1/6$ such that for lower values, i.e., $E \leq E_e=1/6$, the isopotential curves are closed. For larger values $E > E_e=1/6$ the isopotential curves are open and the particles can escape from the scattering region and go to infinity through three different \textit{exits} (See Fig.~\ref{fig:pot}(b)).
\\\indent
Much work has been done in the past decade concerning the effect of
different perturbations such as noise, dissipation and periodic
forcing
\cite{Bernal,Blesa12,Seoane06,Seoane07,Seoane08,Seoane09,Seoane10}
in the escape dynamics of this system under the assumptions of
Newtonian dynamics. Recently, the H\'{e}non-Heiles Hamiltonian has
been also studied in the relativistic case \cite{Bernal17,Bernal18}.
The evolution of the fractal dimension of the scattering function
has been studied in the presence of dissipation, noise and
relativistic corrections; but not in the presence of forcing. This
is one of the main motivations for which we have decided to analyze
the periodically forced H\'{e}non-Heiles system. \\\indent Because
of the chaotic dynamics, particles with slightly different initial
conditions can describe completely different trajectories and escape
through different exits. Therefore, chaotic scattering implies some
sort of unpredictability, understanding it as the difficulty of
predicting the exit through which a trajectory will escape. In a
Hamiltonian system, since the total energy is conserved, we cannot
speak about attractors, and thus, we cannot define basins of
attraction \cite{Nusse96b, Ott81}. However, in an open Hamiltonian
system, we can define exit basins \cite{Contopoulos02} in an
analogous way to the basins of attraction in dissipative systems. We
say that the exit basin associated with the exit $i$ is the set of
initial conditions whose trajectories will escape through the exit
$i$. In the case of the H\'{e}non-Heiles system, there exist three
exit basins and, in the nonhyperbolic regime, a set of initial
conditions for the particles that will never escape from the
scattering region. To quantify the unpredictability of the escape
basins, we use the basin entropy as a useful tool to analyze the
exit basins associated with a large set of parameters (in our
particular case the forcing amplitude and frequency), which is
another issue we add in this paper.
\\\indent This paper is organized as follows. In Sec.~\ref{sec:Model
Description}, we describe our model, the periodically forced
H\'{e}non-Heiles system. The effects of forcing in the fractal
dimension of the scattering function are carried out in
Sec.~\ref{sec:Fractal Dimension}. In that section we also provide a
theoretical reasoning that could justify the obtained results. The
qualitative effects of the forcing term on the basin topology are
shown in Sec.~\ref{sec:Basin Topology}. In Sec.~\ref{sec: Basin
Entropy}, we evaluate the unpredictability of the exit basins using
the concept of basin entropy. Finally, in Sec.~\ref{sec:
Conclusions}, we present the main conclusions of this manuscript.

%%%%%%%%%%%%%%%%%%%%%%%%%%%%%%%%%%%%%%%%%%%%%%%%%%%%%%%%%%%%%%%%%%%%%%%%%%%%%%%%%%%%%%%%%%%%%%%%%%%%%%%%%%%%%%%%%

\section{Model Description} \label{sec:Model Description}
\begin{figure}[htp]
        \centering
        \includegraphics[width=0.45\textwidth,clip]{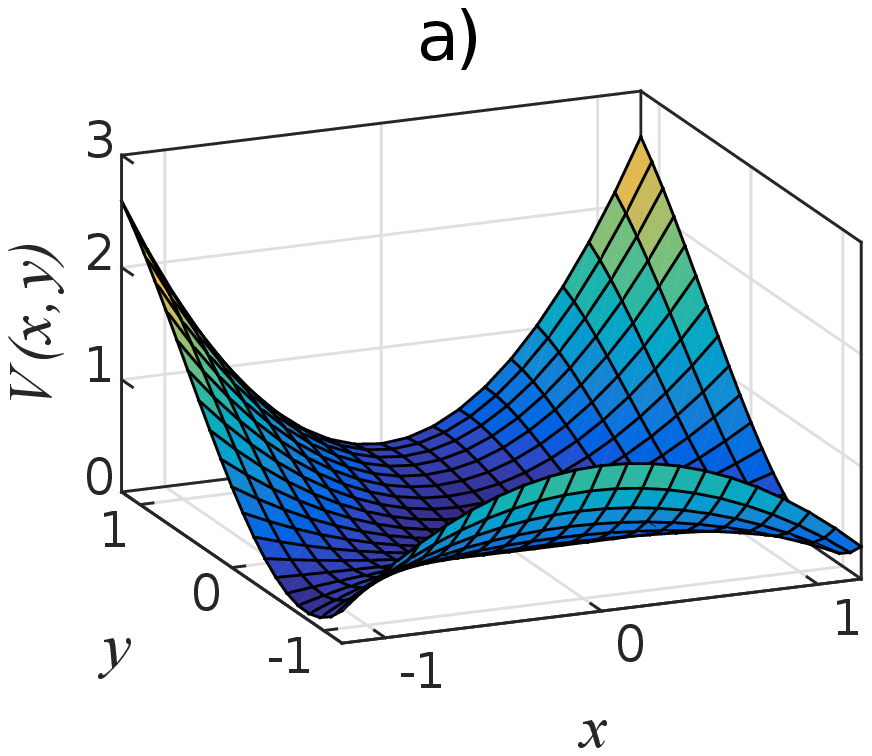} \hspace{.5 cm}
        \includegraphics[width=0.45\textwidth,clip]{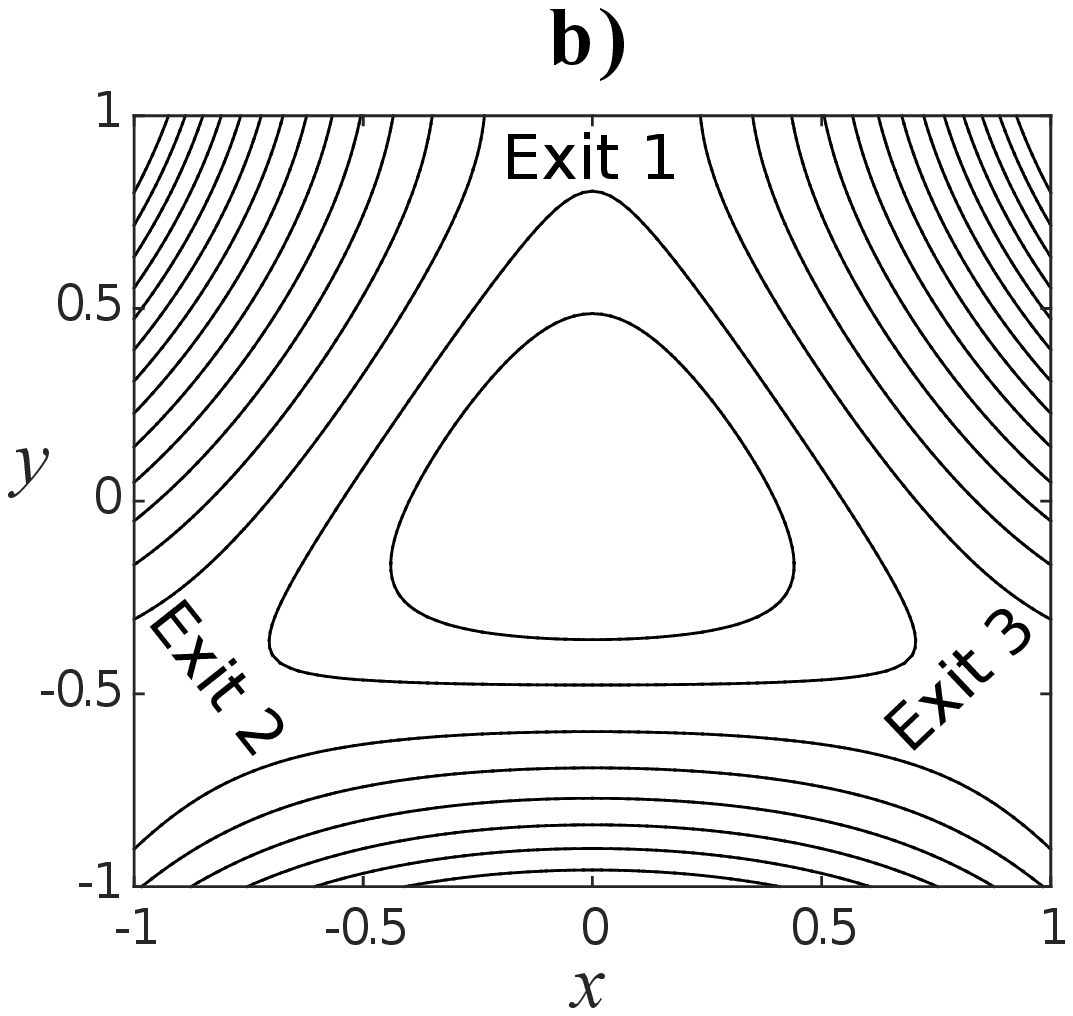}
        \caption{(Color online) (a) Potential energy $V(x,y)$ associated to the the H\'{e}non-Heiles system. (b) The isopotential curves for different values of the energy $E\in[0.08,1.5]$. The curves are closed for energies below the threshold value $E_e=1/6$, and they exhibit three symmetrical exits, separated by an angle of $2\pi/3$ radians, when the energy $E > E_e$.}
        \label{fig:pot}
    \end{figure}

\begin{figure}[htp]
    \centering
    \includegraphics[width=0.98\textwidth,clip]{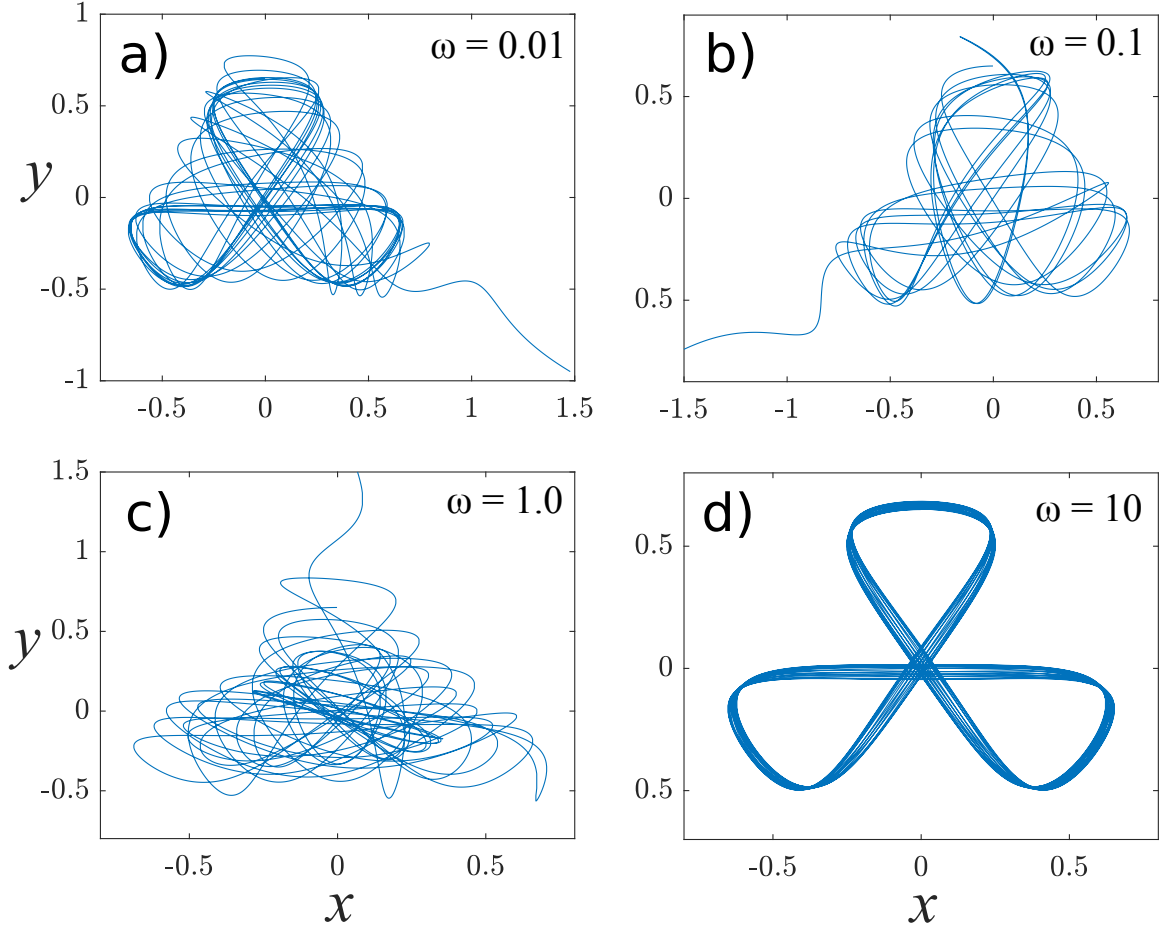}
    \caption{(Color online) Trajectories in the forced H\'{e}non-Heiles system with energy $E=0.19$ and $A=0.01$. Panels (a)-(c) show three trajectories escaping through different exists due to the effects of the periodic forcing term of frequencies $\omega = 0.01$, $\omega = 0.1$ and $\omega = 1.0$, respectively. Panel (d) shows a quasiperiodic orbit for the frequency $\omega = 10.0$. In all the trajectories the initial condition is $(x_0,y_0)=(0.0,0.65)$. This initial condition is associated with a KAM island in the conservative case. We can clearly observe the role of the frequency, which can change drastically the destination of the particle.}  \label{fig:tray12}
\end{figure}

    The H\'{e}non-Heiles system appeared in literature for the first time in 1964 in an article by the astronomers Michel H\'{e}non and Carl Heiles \cite{HH64}. Both researchers worked in the search for a third integral of motion in galactic systems. The equations of motion are given by
    \begin{equation} \label{eq:eq motion}
    \begin{aligned}
    \dot{x} & = {p}, \\
    \dot{p} & = -x - 2xy \\
    \dot{y} & = {q}, \\
    \dot{q} &= -y - x^2 + y^2,
    \end{aligned}
    \end{equation}
\noindent where \textit{p} and \textit{q} denote the two components of the generalized momentum.
\\\indent We focus our attention in the effects of the periodic forcing on the chaotic scattering.
In this context, the periodic forcing is introduced in a natural way as follows \cite{Blesa14}:
\begin{equation} \label{eq:eq motionF}
    \begin{aligned}
        \dot{p} & = -x - 2xy +A_x\sin{\omega_x t},\\
        \dot{q} & = -y - x^2 +y^2+A_y\sin{\omega_y t},
    \end{aligned}
\end{equation}
where we take, for simplicity and without loss of generality, the
same amplitudes ($A_x=A_y=A$) and frequencies ($\omega_x = \omega_y
=\omega$). One of the physical motivations to introduce the forcing
in this kind of systems is the study of spiral galaxies in Astronomy
and Astrophysics in which the forcing is a natural ingredient as
shown in Ref.~\cite{Kandrup04}.
\\\indent
To intuitively visualize the system, we plot the potential and the isopotential curves for different values of the energy in Fig.~\ref{fig:pot}.
\\\indent For energies $E>E_e$ the trajectories may come from infinity or from inside the scattering region, and after interacting with the potential, they escape through one of the exits. One of the properties of the Hénon-Heiles system is the existence of three highly unstable periodic orbits known as Lyapunov orbits \cite{Contopoulos90}, each one located in the vicinity of one of the exits. When a trajectory passes through a Lyapunov orbit with its velocity vector pointing outwards of the scattering region, it will escape to infinity and will never come back.
\\\indent The addition of a periodic forcing term can change the dynamics of the system in a drastic way \cite{Blesa12}. In Fig.~\ref{fig:tray12}, we show four trajectories in the presence of forcing of amplitude $A=0.01$ with different frequencies. Figures~\ref{fig:tray12}(a)-(c) show three trajectories escaping through different exits after passing through the scattering region. In Fig.~\ref{fig:tray12}(d) we show a quasiperiodic orbit. All the trajectories have been launched from the same initial point $ (x_0, y_0) = (0.0,0.65) $ with the same energy $E = 0.19$. In this figure we can see how the frequency of a forcing term with a small amplitude can modify the exit through which the particle escapes.
\\\indent
Even for values $ E > E_e $ there exist initial conditions that generate trajectories that will remain forever within the scattering region. These trajectories are typically quasiperiodic. Quasiperiodic orbits are trajectories that periodically return to a finite region of the phase space but never close on themselves, which belong to a Kolmogorov-Arnold-Moser (KAM)-island \cite{Ott}. The existence of these islands is one of the main characteristics of nonhyperbolic chaotic scattering and as a consequence the decay law of the survival probability is algebraic. When the scattering is hyperbolic, the stable and unstable manifolds of the chaotic saddle are never tangent and every saddle point is hyperbolic \cite{Aguirre09}. In addition there are no KAM islands mixed with the chaotic saddle and consequently the decay law becomes exponential. We are particularly interested among other things in the dynamics of the system associated with KAM islands. This justifies that we have considered the nonhyperbolic regime, which is manifested for energies in the approximate range $ E \in [E_e, 0.23] $ \cite{Blesa12}, while the hyperbolic regime is associated to values of $E \gtrsim 0.23$.

%%%%%%%%%%%%%%%%%%%%%%%%%%%%%%%%%%%%%%%%%%%%%%%%%%%%%%%%%%%%%%%%%%%%%%%%%%%%%%%%%%%%%%%%%%%%%%%%%%%%%%%%%%%%%%%%%%%%%%%%%%%%%%%%%
\section{Fractal dimension}\label{sec:Fractal Dimension}
\begin{figure}[htp]
    \centering
    \includegraphics[width=0.45\textwidth,clip]{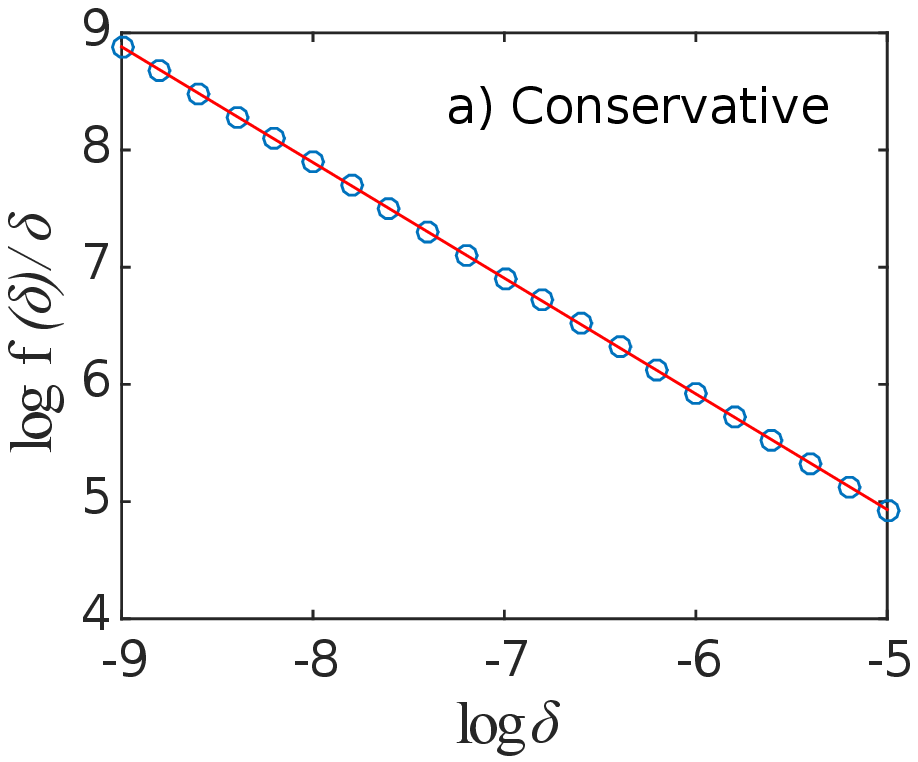}\hspace{.5 cm}
    \includegraphics[width=0.45\textwidth,clip]{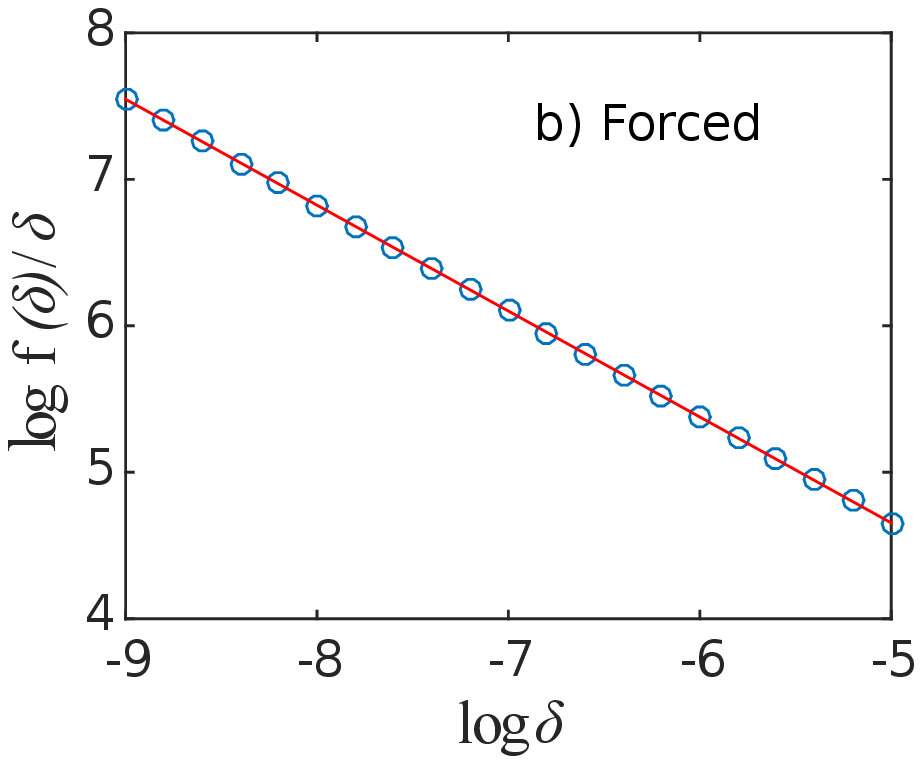}
    \caption{(Color online) Algebraic scaling between $f(\delta)/\delta$ and $\delta$ for $E = 0.17$. (a) For the conservative case. (b) Forced case with amplitude $A = 0.05$ and frequency $\omega = 1$. The fractal dimension is estimated to be $D=0.99$ in the conservative case and $D=0.72$ in the forced case.}  \label{fig:regd}
\end{figure}

\begin{figure}[htp]
    \centering
    \includegraphics[width=0.6\textwidth,clip]{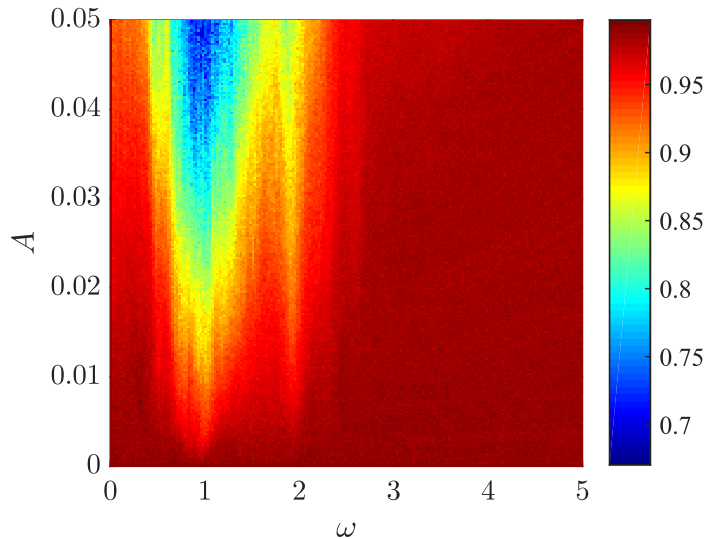}\hspace{.5 cm}
    \caption{(Color online) Color-code map of the fractal dimension for several values of the forcing frequency and amplitude in the periodically forced H\'{e}non-Heiles system with $E=0.17$. We have used $250\times 250$ equally spaced values of the parameters. The hot colors indicate larger values of the fractal dimension. It can be observed that, for any non-zero amplitude, the fractal dimension exhibits a resonant-like evolution, where $ \omega \approx1 $ and $ \omega \approx 2 $ are the main frequencies.}  \label{fig:dAw}
 \end{figure}

The scattering function presents fractal geometry, which implies some sort of unpredictability when relating the input and output variables in the scattering region. We can quantify this unpredictability through the computation of the fractal dimension of the set of singularities of the scattering function. Previous researchers \cite{Lau91} conjectured, providing numerical evidence, that in nonhyperbolic chaotic scattering the set of singularities of the scattering function has Lebesgue measure zero and its fractal dimension is always $ D = 1 $. This value implies that the difficulty to determine the output variable from the input variable is maximal. The value $ D = 1 $ is justified based on the algebraic decay law of the survival probability. \\\indent
To perform the calculation of the fractal dimension, we use the uncertainty algorithm given in \cite{Grebogi83, McDonald85}. In particular, we launch the trajectories from the line segment defined by the points $(x,y)=(0,-0.5)$ and $(x,y)=(0,0)$. In order to fix the value of $\dot{y}$ we use the tangential shooting method \cite{Aguirre01}, in which the trajectories are launched towards the scattering region in such a way that the velocity vector is tangent to a circumference centered at the origin and passing through the point $ (x_0, y_0 )$. For a given initial condition $y_0$ we choose another initial condition $y_0+\delta$, where $\delta$ is some small perturbation. For a fixed value of the uncertainty $\delta$, we compute the escape time $(T)$ of both trajectories, and we say that the initial condition is uncertain if $|T(y_0)-T(y_0+\delta)|>h$, where $h$ is the integration step of the numerical method (a fourth-order Runge-Kutta method in our case).
  \\\indent
 The fraction of uncertain initial conditions obeys the law \cite{Lau91}:
 \begin{equation} \label{eq:f}
 f(\delta) \sim \delta^{1-D}.
 \end{equation}
Taking logarithms in the above equation we obtain
 \begin{equation} \label{eq:logf}
 \log\frac{f(\delta)}{\delta} = -D\log{\delta} +  k,
 \end{equation}
 where $k$ is a constant. This equation allows us to obtain the fractal dimension $D$ computationally from the slope of the line that must yield a representation $ \log{\left(f (\delta) / \delta\right)} $ versus $ \log {\delta} $. \\\indent
 In all the simulations of this section we have set $h=0.005$ and we have taken $21$ values of $\delta$ from $10^{-9}$ to $10^{-5}$. In Fig.~\ref{fig:regd} we represent the results for both the conservative case and forced case with amplitude $A=0.05$ and a frequency $\omega=1$. In this case, in order to obtain the fraction of uncertain initial conditions, we have taken $50000$ initial conditions for each $\delta$. In both cases we observe a strict linear relation between the variables. The fractal dimension is estimated to be $D=0.99$ in the conservative case and $D=0.72$ in the forced case.
\\\indent First, we want to clarify the effect of the frequency and the amplitude on the fractal dimension. We have obtained the fractal dimension for $ 250 \times 250 $ values of $ A \in [0,0.05] $ and $ \omega \in [0,5] $ for $ E = 0.17 $. In each case the fraction of uncertain initial conditions was obtained by throwing $50000$ initial conditions selected by sweeping along the segment defined by the points $ (x, y) = (0, -0.5) $ and $ (x, y) = (0,0) $. A good way to represent these results is using a color-code map in the $ (\omega, A) $ plane. The results are shown in Fig.~\ref{fig:dAw}. The hot colors indicate larger values of the fractal dimension. In the figure we observe a resonant-like behavior where the critical values for the frequency are $\omega \approx 1$ and $\omega \approx 2$. For values close to $ \omega = 1 $ the fractal dimension drastically decreases and does it again, less abruptly, when $ \omega \approx 2 $. When the second resonant frequency is exceeded, the fractal dimension returns monotonously to $ D \approx 1 $. The decrease occurs for the same value of $ \omega $ regardless of the amplitude, although the decrease will be greater for higher amplitude values.   \\\indent

\begin{figure}[htp]
    \centering
    \includegraphics[width=0.6\textwidth,clip]{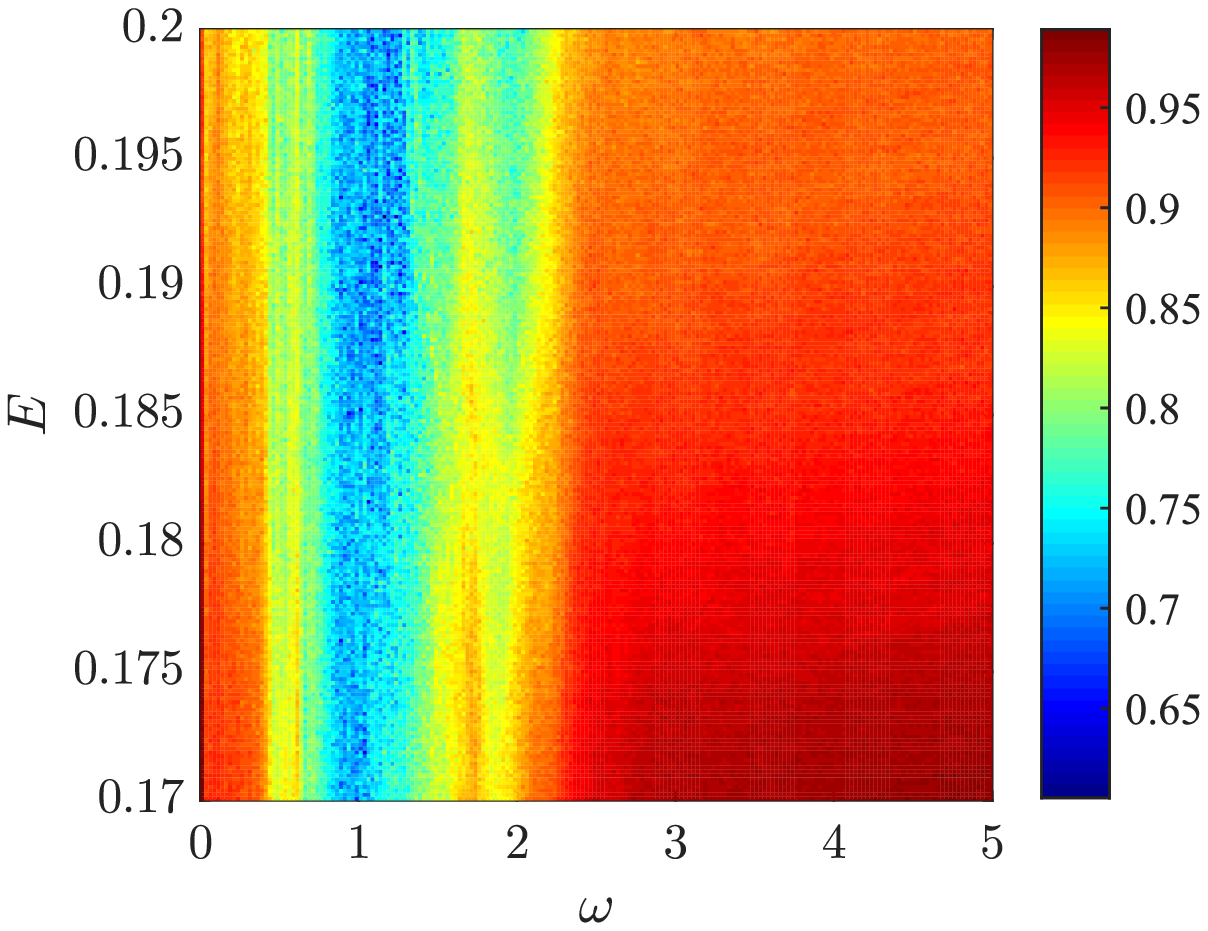}\hspace{.5 cm}
    \caption{(Color online) Color-code map of the fractal dimension for several values of the forcing frequency and energy in the periodically forced H\'{e}non-Heiles system with $A=0.05$. We have used $250\times 250$ equally spaced values of the parameters. The hot colors indicate larger values of the fractal dimension. It can be observed that, for any energy within the nonhyperbolic regime, the fractal dimension exhibits a resonant-like evolution, where $ \omega \approx1 $ and $ \omega \approx 2 $ are the main frequencies.}  \label{fig:dEw}
  \end{figure}

   In order to generalize the previous results for any energy within the nonhyperbolic regime, we have set $ A = 0.05 $ and computed the fractal dimension for different values of the frequency and the energy. Specifically, we have taken $ 250 \times 250 $ combinations of energy $ E \in [0.17,0.20] $ and frequency $ \omega \in [0,5] $. We have again constructed a color-code map, which is represented in Fig.~\ref{fig:dEw}. From this figure it follows that the evolution of the fractal dimension with the forcing frequency is qualitatively the same within the nonhyperbolic regime. In the same way as in Fig.~\ref{fig:dAw}, the minimum of the fractal dimension associated with the resonant frequencies can be clearly observed.      \\\indent

\begin{figure}[htp]
            \centering
            \includegraphics[trim = 10mm 0mm 0mm 0mm, clip,width=0.6\textwidth,clip]{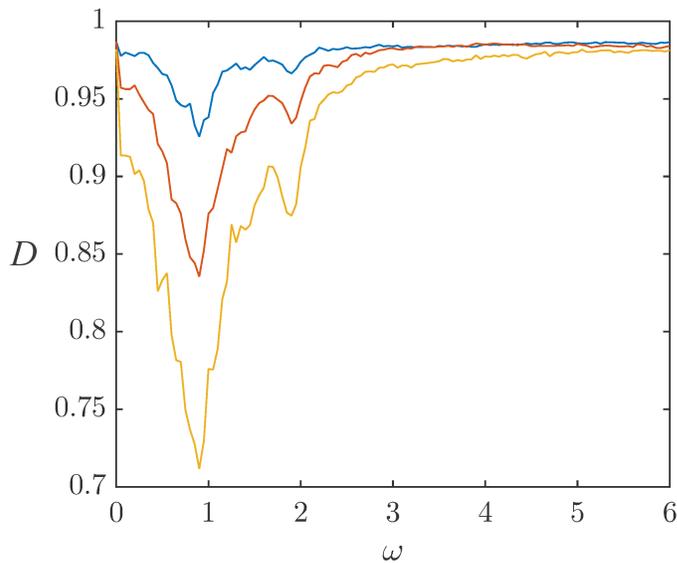}\hspace{.5 cm}
            \caption{(Color online) Fractal dimension of the scattering function for $E=0.17$ and different values of amplitude $A=0.01$ (blue), $A=0.03$ (red) and $A=0.05$ (yellow). For each frequency, to calculate the fraction of uncertain initial conditions, $50000$ initial conditions $\theta_0 \in[0,2\pi]$ have been launched from the point $(x_0,y_0)=(0.15, -0.25)$.  }  \label{fig:dw}
        \end{figure}

    Now we provide a theoretical reasoning that could justify the effects of the forcing near the resonance, using the fractal dimension of Cantor-like structures as a parallelism. In systems where chaotic scattering occurs, particles are launched from a line segment straddling the stable manifold of the chaotic saddle. There is a certain interval of the input variable which lead to trajectories that remain in the scattering region for at least a time $T_0$. By a time $2T_0$ a fraction $\eta$ of the remaining particles leave the scatterer. If these particles are located in the middle of the original interval, we are left with two equal-length subintervals of the input variable that lead to trajectories that do not escape for, at least, a duration of time $2T_0$. By a time $3T_0$ an additional fraction $\eta$ of the particles remaining at time $2T_0$ leave the scatterer. We assume that these particles were located in the middle of the first two subintervals. If we continue this iterative proceeding we obtain a Cantor-like set of Lebesgue measure zero with associated fractal dimension, $D$, given by
    \begin{equation} \label{eq:fd}
        D = \frac{\ln{2}}{\ln{[2/(1-\eta)]}} \cdot
    \end{equation}
    \\\indent On the other hand, for high amplitudes, near the resonant frequency, say $\omega \in[0.8,1.2]$, KAM islands are destroyed and then the decay law becomes exponential. In this case the decay rate is related to the fraction $\eta$ remaining at each stage of the construction of the Cantor-like set by
    \begin{equation} \label{eq:fd2}
        \gamma(\omega) = \frac{1}{T_0}\ln{(1-\eta)^{-1}}\cdot
    \end{equation}
    \\\indent According to Eqs.~(\ref{eq:fd}) and (\ref{eq:fd2}) we find a relation between the fractal dimension and the decay rate
    \begin{equation} \label{eq:fd3}
        D = \frac{\ln 2}{\ln{2}+T_0\gamma(\omega)}\cdot
    \end{equation}
    \\\indent If we increase $\omega$, approaching the resonant frequency, the amount of particles remaining at each time is reduced and then we expect $d\gamma/d \omega >0$. On the other hand, once the resonant frequency is reached, if we increase $\omega$ we expect $d\gamma/d \omega <0$. According to Eq.~\ref{eq:fd3}, since $\gamma(\omega)>0$, $d\gamma/d \omega >0$ implies $dD/d \omega <0$. In the same way $d\gamma/d \omega <0$ leads to $dD/d \omega >0$. These theoretical reasoning is in good agreement with the numerical results showed in Fig.~\ref{fig:dAw} and Fig.~\ref{fig:dEw}. In order to observe clearly the change in sign of $dD/d \omega$ in Fig.~\ref{fig:dw}, we plot the computed fractal dimension of the scattering function versus the forcing frequency for $E=0.17$ and different values of the amplitude. To perform the dimension calculation we launched, for each frequency, $50000$ initial conditions $\theta_0 \in[0,2\pi]$ from the point $(x_0,y_0)=(0.15, -0.25)$. In this figure, we clearly observe the decrease of the fractal dimension near the frequencies $\omega\approx 1$ and $\omega \approx 2$. The resonant-like behavior occurs for the three amplitudes considered, being the decrease of the fractal dimension greater for larger amplitudes.

%%%%%%%%%%%%%%%%%%%%%%%%%%%%%%%%%%%%%%%%%%%%%%%%%%%%%%%%%%%%%%%%%%%%%%%%%%%%%%%%%%%%%%%%%%%%%%%%%%%%%%%%%%%%%%%%%

\section{Basin topology}\label{sec:Basin Topology}
\begin{figure}[htp]
    \centering
    \includegraphics[width=0.48\textwidth,clip]{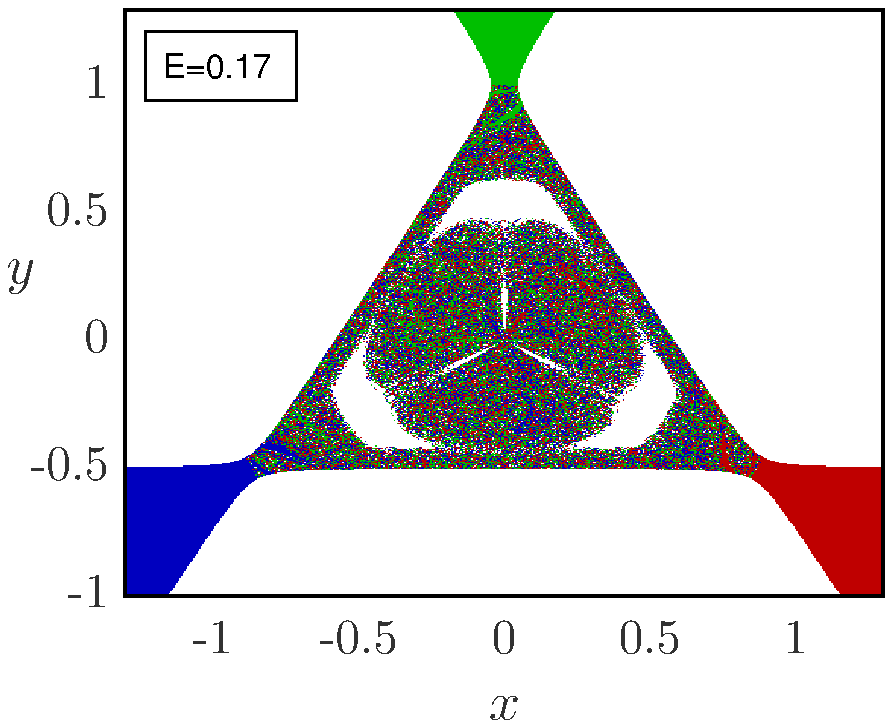}
    \includegraphics[width=0.48\textwidth,clip]{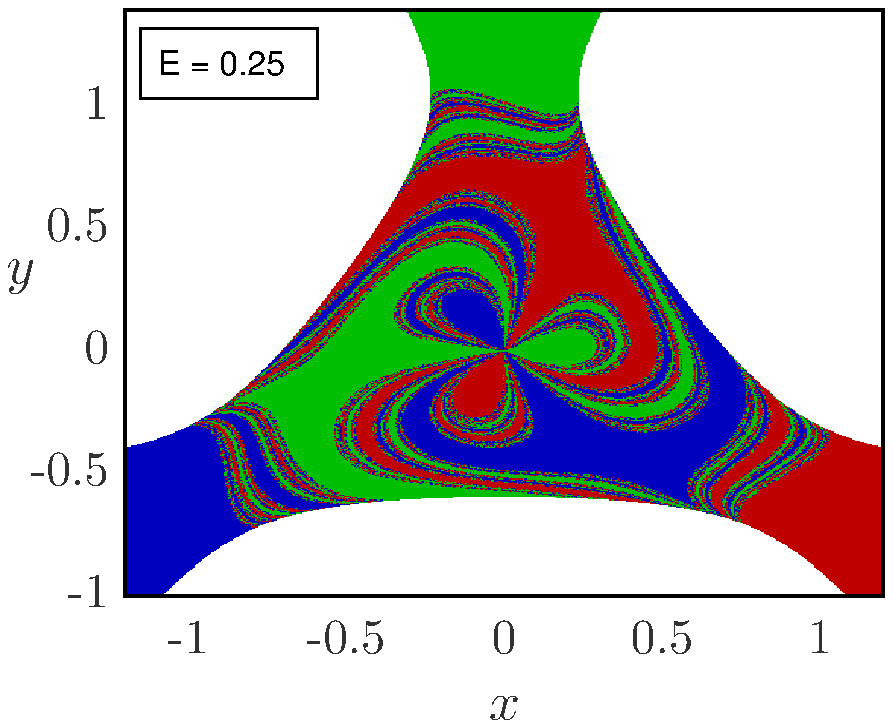}
    \caption{(Color online) Exit basins in the physical space $(x,y)$ for the conservative H\'{e}non-Heiles system for the energy values indicated in each figure. The color code is as follows: green, blue and red correspond to the initial conditions that lead to exits 1, 2 and 3, respectively, and white corresponds to bounded orbits that never escape.}  \label{fig:basins1}
\end{figure}

\begin{figure}[htp]
    \centering
    \includegraphics[trim = 0mm 0mm 0mm 0mm, clip,width=1\textwidth]{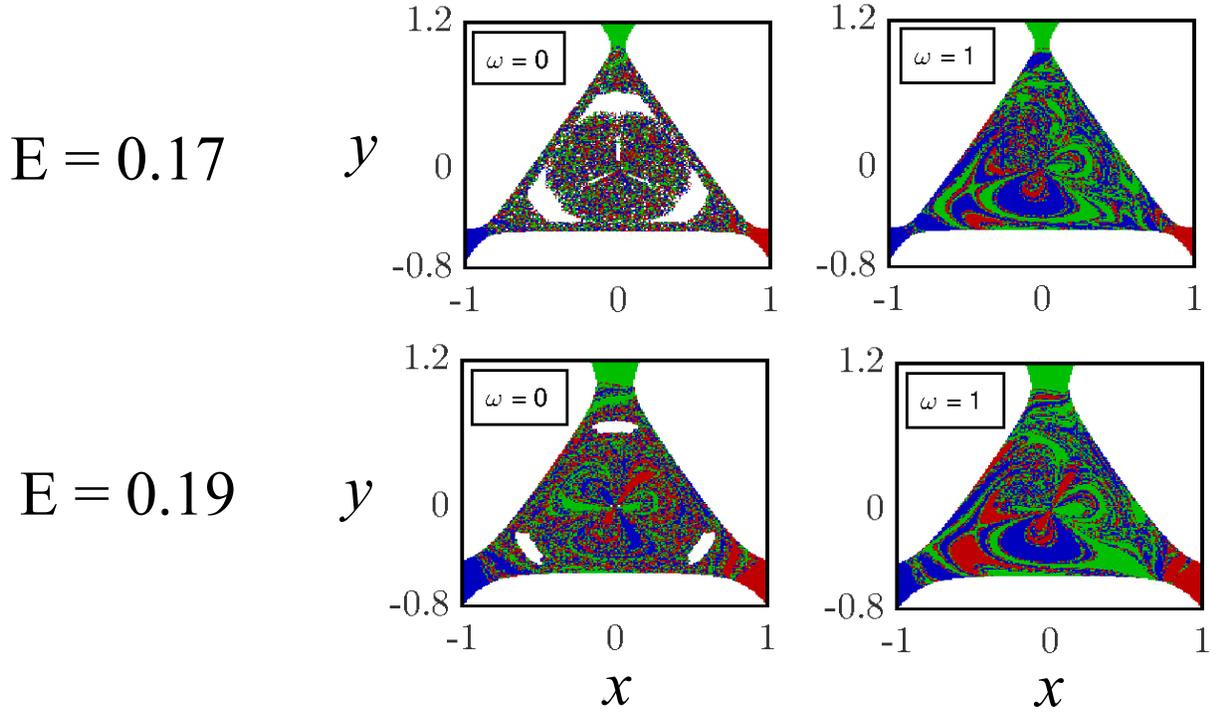}\hspace{.5 cm}
    \caption{(Color online) Exit basins in the physical space $(x,y)$ for the forced H\'{e}non-Heiles system when the forcing amplitude is $A=0.05$. The energy value is $E=0.17$ (upper) and $E=0.19$ (lower) and the frequency value is indicated in each figure. The color code is as described in the caption to Fig.~\ref{fig:basins1}.}  \label{fig:KAMdestroy}
\end{figure}

\begin{figure}[htp]
        \centering
        \includegraphics[width=0.5\textwidth,clip]{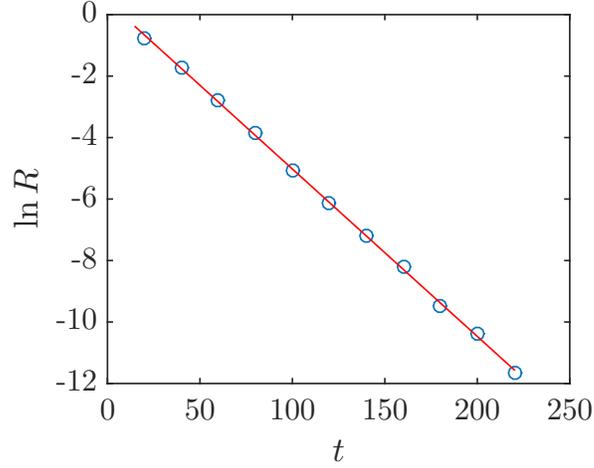}\hspace{.5 cm}
        \caption{(Color online) Exponential decay law for the particles remaining in the scattering region at time $t$. $R$ denotes the fraction of particles that survive in a time $t$. The energy is $E=0.19$ and the forcing amplitude and frequency are $A=0.05$ and $\omega = 0.9$. The decay rate is estimated to be $\gamma = 5.45\times10^{-2}$.}  \label{fig:decay}
    \end{figure}

\begin{figure}[htp]
        \centering
        \includegraphics[width=1\textwidth,clip]{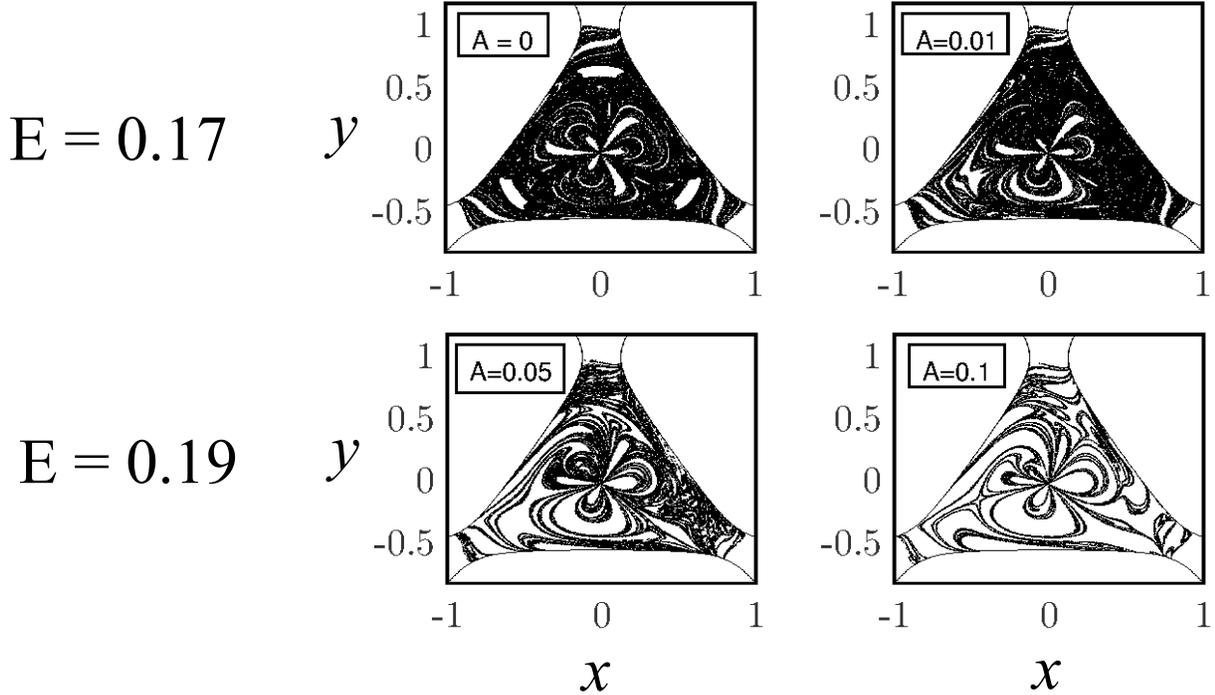}\hspace{.5 cm}
        \caption{Boundaries of the exit basins in the physical space $(x,y)$ when the energy is $E=0.19$ and the forcing frequency $\omega=0.9$. The value of the amplitude is indicated in each figure. The fraction occupied by the boundaries is:  0.82 ($A=0$), 0.71 ($A=0.01$), 0.45 ($A=0.05$) and 0.23 ($A=0.1$). }  \label{fig:boundaries}
     \end{figure}

To obtain the exit basins, we choose a uniform grid of $ 500 \times 500 $ initial conditions in the plotted region. We plot each initial condition with a different color depending on the exit  through which the trajectory escapes. We have selected the initial conditions in the physical space $ (x, y) $ using the tangential shooting method. In Fig.~\ref{fig:basins1} we represent two exit basins in the conservative case with different values of the energy. In the basin obtained for $E=0.17$ we can observe well-defined regions where the particles will never escape (nonhyperbolic regime), but the same does not occur for $E=0.25$ (hyperbolic regime).\\\indent
\\\indent The destruction of the KAM islands can be observed intuitively in the exit basins. For two different energies in the nonhyperbolic regime ($ E = 0.17 $ and $ E = 0.19 $) we have obtained the exit basins with resolution $ 500 \times 500 $ for a forcing amplitude $ A = 0.05 $ and $51$ frequencies in the range $\omega\in[0,5]$. The only frequencies that lead to the destruction of the KAM islands are $ \omega = 0.9 $ and $ \omega = 1.0 $. In  Fig.~\ref{fig:KAMdestroy} we plot the exit basins for $\omega = 0.0 $ and $\omega = 1.0 $.
    \\\indent A natural question at this point is: what is the critical value of the forcing amplitude, $A_c$, that allows the destruction of the KAM islands in the resonance? We have obtained its value for $14$ values of the energy in the range $E\in[0.17,0.2]$ for $\omega=0.9$. The value obtained is $A_c = 0.015$ in all cases.
         \\\indent In nonhyperbolic chaotic scattering the particle decay law is algebraic and there exist KAM islands mixed with the chaotic saddle in the phase space. But if the amplitude is large enough, a forcing close to the resonance frequency $ \omega = 1 $ destroys the KAM islands. Therefore all the particles will escape in a finite time and the decay law becomes exponential
 \begin{equation} \label{eq:decayexp}
 R(t) \sim e^{-\gamma t},
 \end{equation}
 where $ R (t) $ is the fraction of particles that survive at time $ t $ and $ \gamma \geq 0 $ is the decay rate.\\\indent
For values $ \omega \approx 1 $ the decay rate is approximately constant within the nonhyperbolic regime. For 31 values of the energy in the range $ E \in [0.17, 0.2] $ we have obtained (with $A=0.05$) $\gamma = (3.318\pm 0.033)\times 10^{-2}$ for $\omega = 1.0$ and    $\gamma = (5.648\pm0.097)\times 10^{-2}$ for $\omega = 0.9$. We have considered an error $\varepsilon_\gamma = 3\sigma(\gamma)/\sqrt{n} $, where $n=31$ is the number of samples and $\sigma$ is the standard deviation. To obtain the decay rate we have used $10^6$ initial conditions, using an equally spaced $1000\times1000$ grid in the physical space $(x, y)$. For each initial condition we compute the escape time and determine the fraction that remains in steps $ \Delta t = 20 $. Finally, we represent $ \ln R $ versus $ t $ and we obtain the decay rate from the slope of the straight line. For example, Fig.~\ref{fig:decay} shows the straight line obtained by the least squares method for $ E = 0.19 $ with a frequency forcing $ \omega = 0.9 $ and amplitude $ A = 0.05 $.
    \\\indent Another consequence of a forcing with a frequency close to resonance is the decrease in the area occupied by the basin boundaries. Given an energy $E=0.19$ and a forcing frequency $\omega=0.9$, the fraction of the area of the basins occupied by the boundaries decreases with increasing amplitude, as shown in Fig.~\ref{fig:boundaries}, where we plot the boundaries of the basins for different values of the amplitude. The fraction occupied by the boundaries is:  0.82 ($A=0$), 0.71 ($A=0.01$), 0.45 ($A=0.05$) and 0.23 ($A=0.1$).

%%%%%%%%%%%%%%%%%%%%%%%%%%%%%%%%%%%%%%%%%%%%%%%%%%%%%%%%%%%%%%%%%%%%%%%%%%%%%%%%%%%%%%%%%%%%%%%%%%%%%%%%%%%%%%%%%%%%%%%%%
\section{Basin Entropy} \label{sec: Basin Entropy}
\begin{figure}[htp]
            \centering
            \includegraphics[width=0.6\textwidth,clip]{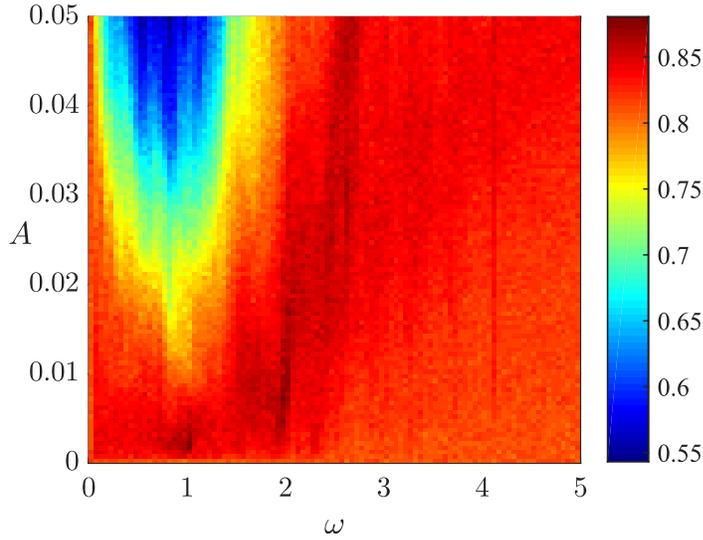}\hspace{.5 cm}
            \caption{(Color online) Color-code map of the basin entropy for several values of the forcing frequency and amplitude, with $E=0.19$. We have used $100\times 100$ values of the parameters. The hot colors indicate larger values of the basin entropy.}  \label{fig:sbcm}
        \end{figure}

\begin{figure}[htp]
        \centering
        \includegraphics[width=0.6\textwidth,clip]{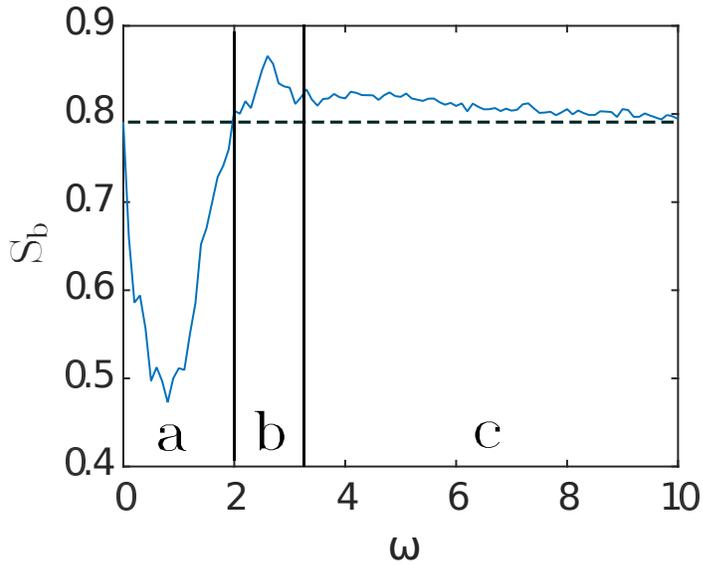}\hspace{.5 cm}
        \caption{(Color online) Evolution of the basin entropy with the frequency of the periodic forcing for $ E = 0.19 $ and $ A = 0.05 $. The dashed line represents the value of basin entropy of the undisturbed case. Three different regions are represented: (a) $ \omega \in (0, 2.0) $, (b) $\omega \in (2.0,3.2)$ and (c) $\omega \in (3.2,10.0)$. It can be seen the resonant-like behavior at $\omega \approx 1$ and a maximum in the basin entropy for $\omega \approx 2.5$.}  \label{fig:sbw}
    \end{figure}

Sometimes in nonlinear dynamics, unpredictability is defined as the difficulty to predict the evolution of the orbits \cite{Adler,Hunt,Sinai}. In the context of open Hamiltonian systems, we consider unpredictability as the difficulty in determining the final state of a system from certain initial conditions. In this sense, the topology of the exit basins in open Hamiltonian systems, or the topology of the basins of attraction in dissipative systems, is closely related to the unpredictability of the system. One probably say that in Fig.~\ref{fig:basins1} the basin for $E=0.17$ is more unpredictable than the basin for $E=0.25$. The basin entropy \cite{Daza16} is a tool that allows us to quantify the unpredictability that we detect intuitively in the basins. It also allows to study and quantify the unpredictability for a large set of basins.   \\\indent
The method to compute the basin entropy is as follows. We randomly select $N$ overlapping square boxes of linear size $\varepsilon$ and we obtain the entropy of each one of the boxes:
    \begin{equation} \label{eq:sb1}
    S_i = \sum_{j=1}^{m_i}\frac{n_{i,j}}{\varepsilon^2}\ln\left(\frac{\varepsilon^2}{n_{i,j}}\right),
    \end{equation}
    where $m_i$ is the number of different destinations (\textit{colors}) in the box $i$ and $n_{i,j}$ is the number of points with color $j$ in the box. In the H\'{e}non-Heiles system $m_i \in[1,3]$ in the hyperbolic regime and $m_i \in[1,4]$ in the nonhyperbolic regimen, due to the existence of quasiperiodic orbits.
    \\\indent The entropy associated to all the $N$ boxes is:
    \begin{equation}\label{eq:sb4}
        S = \sum_{i=1}^{N}\sum_{j=1}^{m_i}\frac{n_{i,j}}{\varepsilon^2}\ln\left(\frac{\varepsilon^2}{n_{i,j}}\right).
        \end{equation}
\\\indent Finally, we obtain the basin entropy by scaling the total entropy $S$ to the total number of boxes, $S_b=S/N$. After this scaling the basin entropy is normalized between $0$ and $\ln N_d$, where $ N_d $ is the number of different destinations. The value $0$ is associated with a basin that has a unique destination and the value $\ln N_d$ is associated with a basin with equiprobable and randomly distributed destinations.
\\\indent In all the simulations of this section we work in the region $\Omega \in[-1,1]\times[-0.8,1.2]$ of the physical space $(x,y)$, using exit basins of resolution $500\times500$. The value of the basin entropy depends on the linear box size $\varepsilon$. Therefore we have to select an adequate value of $\varepsilon$ that allows a reliable portrait of the unpredictability of the basins. We have chosen $\varepsilon = 0.02$, that is, 25 trajectories in each box. We consider that this value accounts for the internal structure of the basins and allows a statistically significant approximation of the probabilities of each color in the box.
\\\indent We have mentioned that some of the effects of a forcing close to the resonant frequency are the destruction of the KAM islands and the decrease in the area occupied by the boundaries of the exit basins. The basin entropy gives an account, among other things, for these two important changes in the topology and allows a reliable portrait of the evolution of unpredictability according to the forcing frequency and amplitude. For $E=0.19$ we have computed $100\times100$ basins for different combinations of amplitude $A\in[0,0.05]$ and frequency $\omega\in[0,5]$. For each of these $10000$ basins we have obtained the basin entropy, shown in the color-code map of Fig.~\ref{fig:sbcm}.
    \\\indent From the color-code map we conclude that there is a resonant-like behavior for the evolution of the unpredictability of the exit basins. The main resonant frequency is $\omega\approx 1$ and it coincides with that obtained in the case of the fractal dimension of the scattering function in Sec.~\ref{sec:Fractal Dimension}.
    \\\indent The color-code map also shows that the basin entropy exceeds the value of the unperturbed case for a frequency value $\omega \approx 2.5$. To observe this point with greater clarity, in Fig.~\ref{fig:sbw} we represent the basin entropy as a function of the forcing frequency for $ E = 0.19 $ and $ A = 0.05 $. In this simulation, 100 exit basins with a resolution $1000\times1000$ have been computed. In Fig.~\ref{fig:sbw} we can distinguish three different regions: (a) $ \omega \in (0, 2.0) $, resonant-like behavior; (b) $\omega \in (2.0,3.2)$, maximum value of basin entropy and (c) $\omega \in (3.2,10.0)$, the forcing becomes unproductive. In the region (a) the basin entropy decreases until reaching its minimum value for $\omega \approx 1.0$ and then increases until reaching the value of the undisturbed case. In the region (b) the maximum value of the basin entropy is obtained for $\omega \approx 2.5$. For this frequency value the fractal dimension of the scattering function was not noticeably affected, but the same does not happen with the basin entropy. This is because this maximum is associated with an increase in the area occupied by the KAM islands. Since in our simulations we have considered that the particles that never escape constitute a destination of the dynamical system, the basin entropy is influenced by both the topology and the area of the KAM islands. The scattering function represents the relation between the escape times of particles and one of the characteristic parameters of the system. Therefore trajectories that never escape are excluded from the scattering function and their fractality is not influenced by the changes in the KAM islands. Finally, in the region (c) the forcing begins to be irrelevant and the basin entropy converges monotonously to the undisturbed case, as shown in the horizontal dashed line. In this last situation, the frequency has no visible effects on the unpredictability of the system when it is large enough, as shown in Ref.~\cite{Blesa14}.

\section{Conclusions } \label{sec: Conclusions}
In summary, our investigations in forced chaotic scattering reveal a resonant-like behavior in the fractal dimension of the scattering function and also in the uncertainty of the exit basins, which has been measured using the the uncertainty algorithm and the basin entropy, respectively. In line with previous works, the main resonant frequency obtained  is $\omega \approx 1.0$, for which both magnitudes are drastically reduced. As the forcing frequency increases, moving away from resonance, the forcing becomes irrelevant and the fractal dimension and the basin entropy return to their value associated with the conservative case. The resonant-like behavior appears for any nonzero amplitude and for any energy value within the nonhyperbolic regime. The decrease in the basin entropy near the main resonant frequency is due, among other things, to the reduction of the area occupied by the KAM islands and the basin boundaries. We have provided theoretical reasoning that could justify the resonant-like behavior in the fractal dimension. These arguments, based on Cantor-like structures, relate the changes in the decay rate of the exponential decay law with the fractal dimension.
\\\indent We have explored the changes in the basin topology and in the escape dynamics in the resonance. We have found the amplitude value, $A_c$, that allows the complete destruction of the KAM islands, which is approximately constant within the nonhyperbolic regime. The same happens with the decay rate of the exponential decay law.
\\\indent A context of physical interest where our work can potentially be useful is in research fields related to laser-driven reactions, chaotic Hamiltonian pumps and oscillations in chemical reactions \cite{Kawai,Hennig,Ramaswamy,Dittrich,Zhang}, among others. In the context of celestial mechanics, in which the H\'{e}non-Heiles system arises, periodic forcing could be used for modeling the effect of companion galaxies \cite{Kandrup04}, such as the Magellanic Clouds orbiting the Milky Way galaxy.

\section*{ACKNOWLEDGMENTS}
This work has been supported by the Spanish Ministry of Economy and Competitiveness and by the Spanish State Research Agency (AEI) and the European Regional Development Fund (FEDER) under Project No. FIS2016-76883-P. M.A.F.S. acknowledges the jointly sponsored financial support by the Fulbright Program and the Spanish Ministry of Education (Program No. FMECD-ST-2016).

\end{document}